\documentclass[12pt,preprint]{aastex}

\def\arcsecpoint{$''\!.$}
\def\deg{$^{\rm o}$}

\def\gtsim{\raisebox{-.5ex}{$\;\stackrel{>}{\sim}\;$}}

\shortauthors{Crenshaw et al.}
\shorttitle{Velocity Offsets in AGN}

\begin{document}

\title{Radial Velocity Offsets Due to Mass Outflows and Extinction in
Active Galactic Nuclei\altaffilmark{1}}

\author{D.M. Crenshaw\altaffilmark{2},
H.R. Schmitt\altaffilmark{3},
S.B. Kraemer\altaffilmark{4},
R.F. Mushotzky\altaffilmark{5}
and J.P. Dunn\altaffilmark{6}}

\altaffiltext{1}{Based on observations made with the NASA/ESA Hubble Space 
Telescope, obtained at the Space Telescope Science Institute, which is 
operated by the Association of Universities for Research in Astronomy,
Inc. under NASA contract NAS 5-26555.}

\altaffiltext{2}{Department of Physics and Astronomy, Georgia State 
University, Astronomy Offices, One Park Place South SE, Suite 700,
Atlanta, GA 30303; crenshaw@chara.gsu.edu}

\altaffiltext{3}{Remote Sensing Division, Naval Research Laboratory,
Washington, DC 20375; and Interferometrics, Inc., Herndon, VA 20171}

\altaffiltext{4}{Institute for Astrophysics and Computational Sciences,
Department of Physics, The Catholic University of America, Washington, DC
20064}

\altaffiltext{5}{NASA/Goddard Space Flight Center, Greenbelt, MD 20771,
USA}

\altaffiltext{6}{Department of Physics, Virginia Tech, Blacksburg, VA
24061}

\begin{abstract}

We present a study of the radial velocity offsets between narrow emission
lines and host galaxy lines (stellar absorption and H~I 21-cm emission) in
Seyfert galaxies with observed redshifts less than 0.043. We find that 35\%
of the Seyferts in the sample show [O~III] emission lines with blueshifts
with respect to their host galaxies exceeding 50 km s$^{-1}$, whereas only
6\% show redshifts this large, in qualitative agreement with most previous
studies. We also find that a greater percentage of Seyfert 1 galaxies show
blueshifts than Seyfert 2 galaxies. Using {\it HST}/STIS spatially-resolved
spectra of the Seyfert 2 galaxy NGC~1068 and the Seyfert 1 galaxy NGC~4151,
we generate geometric models of their narrow-line regions (NLRs) and inner
galactic disks, and show how these models can explain the blueshifted
[O~III] emission lines in collapsed STIS spectra of these two Seyferts. We
conclude that the combination of mass outflow of ionized gas in the NLR and
extinction by dust in the inner disk (primarily in the form of dust
spirals) is primarily responsible for the velocity offsets in Seyfert
galaxies. More exotic explanations are not needed. We discuss the
implications of this result for the velocity offsets found in higher
redshift AGN.

\end{abstract}

~~~~~

\section{Introduction}

Seyfert galaxies, quasars, and other active galactic nuclei (AGN) often
show radial velocity offsets between their broad emission lines, narrow
emission lines, and host galaxy lines (stellar absorption and/or H~I 21-cm
emission), and there is a long history of investigations into the nature of
these offsets, beginning with Penston (1977), Netzer (1977), Osterbrock \&
Cohen (1979), and Gaskell (1982). In addition, many studies have
concentrated on differences in line profile, asymmetry, width (e.g.,
full-width at half maximum [$FWHM$]), peak, and/or velocity centroid as a
function of ionization potential or critical density for lines from the
same region, such as the narrow-line region (NLR; Osterbrock 1981; Pelat,
Alloin, \& Fosbury 1981; De Robertis \& Osterbrock 1986; De Robertis \&
Shaw 1990; Busco \& Steiner 1992; Moore, Cohen, \& Marcy 1996; Komossa et
al. 2008a) or broad-line region (BLR; Wilkes 1984; Crenshaw 1986; Osterbrock
\& Mathews 1986; Corbin 1990; Tytler \& Fan 1992; Sulentic, Marziani, \&
Dultzin-Hacyan 2000; Richards, et al. 2002; Snedden \& Gaskell 2007). Even
for the same emission line from a low-redshift AGN, such as the commonly
used [O~III] $\lambda$5007 line, a perusal of the NASA Extragalactic
Database (NED) demonstrates that the heliocentric redshift in terms of
radial velocity (cz) can differ substantially in excess of the quoted
errors from one study to the next, due to wavelength calibration
uncertainties, different measurement techniques, or aperture effects, for
example. Thus, some care must be taken when determining velocity offsets
between different lines, but there is no doubt from the above references
that the offsets in many sources are real and significant.

Recent investigations have sparked renewed interest in the study of
velocity offsets, because they have been used to indicate the possible
existence of multiple AGN in a galaxy, or an AGN that is displaced with
respect to the gravitational center of a galaxy. The most convincing
evidence comes from galaxies that show active nuclei that are spatially
offset from their optical centers, as well as emission lines that are
offset in radial velocity from the systemic velocities of the host
galaxies. For example, the evidence for multiple AGN in NGC~3341 ($z =$
0.027) appears to be incontrovertible (Barth et al. 2008). Three separate
nuclei can be identified in a Sloan Digital Sky Survey (SDSS) image of this
galaxy. The central nucleus has a LINER/H~II composite spectrum, an offset
nucleus at a distance of 5.1 kpc from the center has a Seyfert 2 spectrum
at a velocity offset of $-$200 km s$^{-1}$, and another offset nucleus at a
distance of 8.4 kpc has a possible LINER spectrum with a similar velocity
offset. Barth et al. suggest that the offset nuclei are due to the fueling
of the supermassive black holes (SMBHs) in two dwarf galaxies that are
merging with the primary disk galaxy.

Another AGN that has generated a lot of interest and speculation is the
quasar SDSS J092712.65+294344.0 ($z =$ 0.71), which shows two sets of
emission lines (one set of narrow lines and one set of broad plus narrow
lines) separated by 2650 km s$^{-1}$ (Komossa, Zhou, \& Lu 2008b). In this
case, the SDSS image shows only an unresolved quasar -- multiple nuclei
have not been detected in groundbased images. Most of the suggested
explanations for the double set of emission lines are based on mergers or
interactions of galaxies and, possibly, their SMBHs. Komossa et al. (2008b)
suggest that the system with broad lines is a recoiling SMBH carrying its
BLR and high-ionization NLR with it and leaving behind the bulk of the
(lower-ionization) NLR. The recoil is due to coalescence of two black holes
which, under the right conditions, can result in anisotropic gravitational
radiation that carries away linear momentum, ejecting the merged SMBH at
velocities up to $\sim$4000 km s$^{-1}$ in the opposite direction
(Campanelli et al. 2007; Baker et al. 2008). Another possible explanation
is that this SDSS quasar contains a binary SMBH, in which the BLR and
high-ionization NLR are due to accretion onto the smaller SMBH, and the
low-ionization NLR is an envelope surrounding the pair of SMBHs (Dotti et
al. 2008; Bogdanovi\'{c}, Eracleous, \& Sigurdsson 2009). Heckman et al.
(2009) suggest a third possibility, which has been used to explain the
properties of the nearby AGN NGC 1275: the narrow redshifted lines are
associated with a small galaxy falling into the center of a rich cluster of
galaxies where it encounters a large galaxy with an AGN that shows both
broad and narrow lines. Finally, Shields, Bonning, and Salviander (2009)
suggest that the double set of emission lines in SDSS J092712.65+294344.0
may be due to a chance alignment of two AGN, possibly within a massive
cluster.

Although two sets of emission lines that are offset in radial velocity are
rare, many AGN show velocity offsets between the emission lines from their
NLRs and the stellar absorption lines or H~I 21-cm emission from their host
galaxies. Comerford et al. (2009) present a study in which 32 of 91 Seyfert
2 galaxies with red host galaxies at 0.34 $< z <$ 0.92 show velocity
offsets between their [O~III] emission lines and stellar absorption lines
that lie in the range $\sim$50 km s$^{-1}$ to $\sim$300 km s$^{-1}$ (two of
these AGN show double peaked emission lines as well). Comerford et al.
suggest that the velocity offsets are likely due to displaced AGN that are
moving with respect to their host galaxies. The same explanations for the
double set of emission lines in SDSS J092712.65+294344.0 can then be
invoked to explain the velocity offsets between the NLR and host galaxy, if
one of the AGN in these scenarios is dispensed with or replaced by an
inactive SMBH. However, Comerford et al. (2009) suggest that the most
plausible explanation is inspiralling SMBHs after a galaxy merger, which
move to the gravitational center of the merger remnant via dynamical
friction. They leave open the possibility, however, that the velocity
centroids of the emission lines are offset from the systemic velocity of
the host galaxy due to a combination of outflows and dust extinction in the
NLR, although they say this is an unlikely explanation for their sample.

In this paper, we show that in low-redshift Seyferts, velocity offsets of
the narrow emission lines with respect to the host galaxies are indeed
due to mass outflows in the NLR and partial extinction of the outflows by
circumnuclear dust. This is not a new idea, as a number of previous studies
suggested that the combination of dust, either inside or between the NLR
clouds, and radial motion, in the form of inflows or outflows, could be the
cause of asymmetries and velocity shifts in the narrow lines (Capriotti,
Foltz, \& Byard 1979; Heckman et al. 1981; Whittle 1985; Dahari \& De
Robertis 1988; De Robertis \& Shaw 1990; Veilleux 1991a). Subsequently, a
number of studies
based on long-slit observations with ground-based telescopes suggested that
outflows and dust obscurations could produce the observed profile
asymmetries (Storchi-Bergmann, Wilson, \& Baldwin 1992;  Arribas,
Mediavilla, \& Garc\'{i}a-Lorenzo 1996; Christopoulou et al. 1997;
Rodr\'{i}guez-Ardila et al. 2006). We can now investigate this
issue in more detail using spatially-resolved optical spectra of the NLR
($\sim$0\arcsecpoint1 resolution) from the Space Telescope Imaging
Spectrograph (STIS) on board the {\it Hubble Space Telescope} ({\it HST}).

\section{Characterization of the Velocity Offsets}

First, we provide a new characterization of the velocity offsets of the
narrow lines in Seyfert galaxies, this time separated by Seyfert type. The
best sample to use is that of Nelson \& Whittle (1995), which is based on
measurements of moderate-resolution (80 -- 230 km s$^{-1}$) ground-based
spectra of a large collection of AGN (mostly Seyfert galaxies) with $z <$
0.043. The advantage of this dataset, in addition to its size and spectral
resolution, is that the observations cover regions in which both strong
emission lines and stellar absorption lines are present, reducing possible
systematic effects. The blue spectra cover H$\beta$ and [O~III]
$\lambda\lambda$4959, 5007 emission and Mg~I $\lambda\lambda\lambda$5167,
5173, 5184 stellar absorption, and the red spectra cover [S~III]
$\lambda$9069 emission and Ca~II $\lambda\lambda\lambda$8498, 8542, 8662
stellar absorption. Nelson \& Whittle show that their stellar velocities,
combined from the Mg~I and Ca~II triplets, are in agreement with the
published H~I 21-cm velocities to within the published uncertainties.

From Table 4 in Nelson \& Whittle (1995), we adopted the stellar absorption
velocity ($v_{abs}$) and used the [O~III] emission velocity centroid
($v_{em}$) for each AGN to determine the velocity offset: $\Delta v =
v_{emis} - v_{abs}$. In the few cases where [O~III] velocities were not
listed, we used [S~III] velocities instead. The mean uncertainties of the
sample are 15 km s$^{-1}$ in $v_{abs}$ and 7 km s$^{-1}$ in $v_{em}$, which
added in quadrature yield an uncertainty in $\Delta v$ of 17 km s$^{-1}$.
To be conservative, we consider an offset $\gtsim$ 3$\sigma$, i.e. $|\Delta
v|$ $\geq$ 50 km s$^{-1}$, to be significant. We identified the AGN in this
dataset that are Seyfert galaxies and determined their types using
information in NED (placing Seyfert 1.8s and 1.9s into the type 2
category). This yielded a total of 65 Seyfert 1 or 2 galaxies with velocity
offsets determined in a uniform manner.

In Figure 1, we plot histograms of the velocity offsets in 50 km s$^{-1}$
bins. The distributions of both Seyfert types peak near zero km s$^{-1}$,
but both show an extended tail on the blueshifted side that goes up to
$-$200 to $-$250 km s$^{-1}$. Combining the two histograms, 23/65 (35\%) of
the Seyferts show blueshifted emission lines with $\Delta v \leq$ $-$50
km s$^{-1}$, whereas 4/65 (6\%) of the Seyferts show redshifted lines with
$\Delta v \geq$ $+$50 km s$^{-1}$. These results are in agreement with
many other studies, which find that the [O~III] lines in low-redshift AGN
often have blue asymmetries (more flux in the blue side of the profile)
and/or blueshifted peaks or centroids with respect to their systemic
velocities, whereas few show red asymmetries or redshifted velocity
offsets (Heckman et al. 1981; Whittle 1985; De Robertis \& Shaw 1990;
Veilleux 1991b; Bian, Yuan, \& Zhao 2005; Boroson 2005; Komossa et al.
2008a). One exception is the study of Vrtilek \& Carleton (1985), which
finds a symmetric distribution of velocity offsets between emission and
absorption lines, but their results depend on only 13 Seyfert galaxies and
low-redshift QSOs.

The Seyfert 1 distribution shown in Figure 1 is shifted to more negative
offset velocities (higher blueshifts, on average) than the Seyfert 2
distribution. A Kolmogorov-Smirnov test on the velocity offsets for the two
types gives only a 2.1\% probability that two samples from the same
population would differ by this much, indicating that the distributions are
indeed different. Although this result is based on only 20 Seyfert 1s and
studies of larger samples are needed to test it, it suggests an additional
constraint on possible explanations or models of the velocity offsets.

\section{{\it HST}/STIS Constraints on the NLR Kinematics and Geometry}

As mentioned in the Introduction, a number of previous studies have
suggested that velocity offsets and/or asymmetries of the narrow emission
lines could be produced by a combination of radial motion and dust
extinction in the NLR. We can test this notion using {\it HST}/STIS
long-slit spectra. We have found that that the kinematics of the NLR are
dominated by radial outflow in the four nearby Seyferts that we have
studied to date: NGC~4151 (Hutchings et al. 1998; Crenshaw et al. 2000; Das
et al. 2005), NGC~1068 (Crenshaw \& Kraemer 2000, Das et al. 2006, 2007;
see also Cecil et al. 2002), Mrk 3 (Ruiz et al. 2001), and Mrk~573 (T.
Fischer et al. in preparation, see also Schlesinger et al. 2009). Our
kinematic models of biconical outflow provide a good match to the overall
flow patterns as a function of distance from the central AGN, despite
significant local variations in the velocity fields. But the main arguments
for radial outflow are independent of our detailed models. The velocities
of the bright NLR clouds (or knots) close to the nucleus reach values that
are too high to be explained by any sort of reasonable gravitational
potential (including contributions from the SMBH, nuclear stellar cluster,
and bulge; Das et al. 2007), which rule out infall or orbital motions as
being dominant. Transverse motions of the gas away from the radio jets
(Axon et al. 1998; Capetti et al. 1999) cannot be dominant, because they
can't explain NLRs with primarily blueshifts from one cone and redshifts
from the other (as in NGC~4151, Crenshaw \& Kraemer 2000). Only radial
outflow can match the kinematics of the Seyfert NLRs that we have studied
to date. Furthermore, radial outflow is consistent with the unified model
of Seyfert galaxies (Antonucci 1993), which postulates that an optically
thick torus outside of the BLR produces a bicone of radiation that ionizes
the NLR and that Seyfert type depends on viewing angle with respect to the
bicone axis. The three Seyfert 2 galaxies studied to date show blueshifts
and redshifts on either side of the central AGN, consistent with the
unified model prediction of a bicone axis near the plane of the sky,
whereas the Seyfert 1.5 galaxy NGC 4151 shows blueshifts on one side and
redshifts on the other side of the central AGN, consistent with an
intermediate viewing angle (Crenshaw et al. 2000). A
number of recent studies based on ground-based observations with
adaptive optics and field spectrographs have also found evidence for
radial outflows in the NLRs of Seyfert galaxies (Barbosa et al. 2009;
Riffel et al. 2009; Stoklasov\'{a} et al. 2009; Storchi-Bergmann et al.
2009).

\subsection{Simulation of Ground-Based Spectra}

In order to compare our results with those from previous ground-based
spectra, which were obtained through apertures that include most or all of
the NLRs in even the closest Seyferts (only a few arcsecs in
extent), we use {\it HST}/STIS long-slit spectra that cover most of the NLR
as well. The two datasets that meet this requirement are the STIS
observations of NGC 1068 (Cecil et al. 2002; Das et al. 2006) and NGC 4151
(Das et al. 2005) obtained at multiple parallel slit positions, as shown in
Figures 2 and 3. Spectra were obtained with the G430M grating through the
52$''$ $\times$ 0\arcsecpoint2 slit at a spectral resolving power of $R =$
$\lambda$/$\Delta\lambda$ $\approx$ 9000 and an angular resolution of
$\sim$0\arcsecpoint1 in the cross-dispersion direction. This procedure
resulted in hundreds of spectra of the H$\beta$ and [O~III]
$\lambda\lambda$4959, 5007 lines along the slits. As shown in Figures 2 and
3, the NLRs are dominated by emission-line knots that are resolved and on
the order of a few tenths of an arcsec in size; we found that each knot has
a unique radial velocity and a fairly large velocity dispersion, on the
order of hundreds of km s$^{-1} FWHM$ (Das et al. 2005, 2006).

To simulate the ground-based spectra, we added together all of the spectra
along each slit position, and then added together the spectra from each
slit (positions 6 and 8 in NGC 1068 were excluded because they overlap the
other slit positions). In Figures 4 and 5, we give these ``collapsed''
spectra in terms of flux as a function of radial velocity of the [O~III]
$\lambda$5007 line, with respect to the rest frame established by
observations of H~I 21-cm emission in each galaxy. The unsmoothed profiles
in Figures 4 and 5 show significant structure with multiple peaks. In fact,
they are virtually identical to the [O~III] profiles from the
high-resolution ground-based spectra of Vrtilek \& Carleton (1985) and
Veilleux (1991c), giving us confidence that any [O~III] emission missed by
the STIS long-slit observations does not significantly alter the profile
structure or shape. From our analysis of the resolved STIS spectra, it is
clear that this structure is due to the superposition of velocity profiles
from the numerous emission-line knots seen in Figures 2 and 3, with
each knot having its own peculiar radial velocity with respect to the
general outflow pattern (Das et al. 2005, 2006).

The [O~III] lines in NGC~1068 are much broader than those in NGC~4151
($FWHM =$ 1450 [$\pm$15] km s$^{-1}$ and 200 [$\pm$10] km s$^{-1}$,
respectively), consistent with the overall higher velocities and
dispersions of the emission-line knots in NGC~1068. The [O~III] lines in
NGC~1068 are blended together in their wings, and we deblended the two
lines by taking the $\lambda$5007 profile, reproducing it at the position
of the $\lambda$4959 line, scaling it by one-third (the ratio of the
transition probabilities of the two lines), subtracting it from the blend,
and repeating the process in an iterative fashion. To simulate
low-resolution ground-based spectra at at $R =$1000, we convolved the
unsmoothed spectra with a Gaussian with a $FWHM =$ 500 km s$^{-1}$
(representing the line-spread function) and rebinned them to one-half of a
resolution element. These ``low-resolution spectra'' are also shown in
Figures 4 and 5.

The [O~III] profiles in Figures 4 and 5 are clearly asymmetric, with more
flux in the blue sides of the profiles. To quantify the velocity offsets,
we measured the velocity centroid of the [O~III] $\lambda$5007 line
(averaged over the entire profile) and its velocity peak (defined as the
centroid averaged over the top 15\% of the [O~III] line). In Table 1, we
give the heliocentric velocities from the H~I 21-cm emission, the
stellar absorption lines, and the [O~III] measurements described above, as
well as the velocity offsets relative to the H~I 21-cm velocity.

From Table 1, it is clear that the H~I 21-cm emission and the stellar
absorption features give the same heliocentric velocities, to within the
measurement errors of $\sim$10 km s$^{-1}$. This gives us confidence that
the systemic velocities of the host galaxies are well established. For the
[O~III] measurements, the velocity centroids are the same to within the
errors in both high and low resolution spectra for each galaxy, whereas the
velocity peaks are discrepant, particularly for NGC~1068. Thus, it is clear
that the velocity centroids provide a much more reliable measure of the
velocity offsets. This would be especially true for lower signal-to-noise
spectra, where random variations could alter the location of the peak
emission significantly.

Looking at the high resolution spectra, the velocity centroids of the
[O~III] emission are blueshifted by $-$160 km s$^{-1}$ in NGC 1068 and
$-$60 km s$^{-1}$ in NGC~4151. These fall inside of the range of values
for Seyfert galaxies shown in Figure 1, and are what we consider to be
significant offsets ($|\Delta v| > 50$ km s$^{-1}$). Interestingly, the
velocity offset for NGC~1068, a Seyfert 2, is greater in magnitude
than that for NGC~4151, a Seyfert 1 galaxy. This is likely due to a
combination of higher velocities and higher extinction in NGC~1068 (Das et
al. 2006, 2006), which we discuss further in the next section.

\subsection{Geometry of the Outflows and Extinction}

To investigate the nature of extinction in the NLR, we first review our
previous studies using {\it HST}/STIS long-slit, low-resolution ($R
\approx$ 1000) UV and optical spectra of NGC~1068 (Kraemer \& Crenshaw
2000) and NGC~4151 (Kraemer et al. 2000). The reddening of the narrow
emission lines in these Seyferts, based on the He~II
$\lambda$1640/$\lambda$4686 ratio, varies between $E(B-V) =$ 0.1 and 0.5 in
a nonuniform fashion across their NLRs. Based on our photoionization
models, the reddening cannot come primarily from dust in the ionized gas,
because the NLR clouds contains either no dust or a reduced dust fraction
(between 10\% and 50\%) compared to the normal dust/gas ratio in the local
ISM of our Galaxy; otherwise, lines from the refractory elements (such as
Mg, Si, and Fe) would be much weaker than observed, and Ly$\alpha$ emission
would be quenched, due to multiple resonant scatterings and eventual
absorption by dust grains. Even the NLR clouds with a 50\% dust fraction
cannot produce the observed reddening, since they have a hydrogen column
$N_H < 10^{21}$ cm$^{-2}$, which would produce a reddening of $E(B-V) <$
0.1 (assuming the Galactic reddening law of Savage \& Mathis 1979 and $N_H
= 5.2 \times 10^{21} E(B-V)$ cm$^{-2}$ from Shull \& van Steenberg 1985).
Thus, the reddening is likely produced in patchy dust that is external to
the NLR clouds. Circumnuclear dust spirals seen in the inner disks of most
Seyfert galaxies (Malkan, Gorjian, \& Tam 1998; Regan \& Mulchaey 1999;
Pogge \& Martini 2002, Martini et al. 2003; Deo et al. 2006) are the likely
source of the reddening; they extend over the same scales as the NLR
(hundreds of parsecs) and the dust spirals would provide the observed
patchiness.

To investigate the effects of extinction on mass outflows in the NLR, we
therefore assume that the reddening arises from dust in the inner disk of
the host galaxy. In order to determine the three-dimensional geometry of
the outflows and extinction, we combined our parameters from the biconical
outflow models with the observed parameters of the host galaxy disks. In
Table 2, we give the five required parameters from Das et al. (2005, 2006):
the position angle (P.A.) of the galaxy's major axis, the inclination ($i$)
of the galactic disk (zero corresponds to face-on), the P.A. of the bicone
axis, the inclination of the bicone axis (zero is in the plane of the sky),
and the half-opening angle (H.O.A.) of the outer edge of the bicone. For
the inclinations, we also indicate which side of the structure is closest
to us.

In Figure 6, we show the geometry of the biconical outflow and the host
galaxy disk for NGC~1068 and NGC~4151. For simplicity, we depict the
midplane of the galaxy and the outer surface of the bicone, even though the
bicone extends to an inner opening angle in our kinematic models. The
left-hand side shows the view from Earth, and the right-hand side shows a
viewpoint in which the bicone axis is in the plane of the sky and the
galactic disk is edge-on, to show the true angle between the bicone and the
disk (see the figure captions for more details). The parallel lines on
either side of the galactic midplane show the dust scale height if we
assume a typical value of $\sim$200 pc for spiral galaxies (Xilouris et al.
1999) and a length of $\sim$800 pc for each bicone along its axis (Das et
al. 2005, 2006).

In the case of NGC~1068, the SW side of the galaxy is closer to us, and the
SW cone therefore experiences more extinction. This is consistent with the
[O~III] image in Figure 2, which shows weaker, and in some place absent,
emission SW of the location of the SMBH, which is 0\arcsecpoint15 south of
the ``hot spot'' in Figure 2 (Das et al. 2006). The bicone axis lies nearly
in the plane of the sky and the nearer side of each cone is blueshifted,
whereas the farther side of each cone is redshifted. Because the disk has a
finite thickness, the blueshifted side of each cone in NGC~1068 experiences
less extinction, on average, than the redshifted side of each cone. In the
case of NGC~4151, the SW cone is entirely blueshifted and the NE cone is
entirely redshifted. The host galaxy disk is close to the plane of the sky,
and the SW cone shows less extinction than the NE cone shows. Thus, both
geometries lead to emission-line profiles that are asymmetric to the blue,
in agreement with the observed profiles.

To determine if this explanation makes quantitative sense, we can look at
the asymmetry of the emission-line profiles, and ask if there is enough
dust in the inner galactic disks to produce the asymmetries. For
simplicity, let us assume that the blue wings of the profiles are not
affected by extinction and that the red wings are affected by a simple
screen. In this case, the ratios of blue-to-red wing fluxes for [O~III]
$\lambda$5007 are 1.59 for NGC~1068 and 1.10 for NGC~4151, which yield
extinctions in the V-band of $A_V =$ 0.5 mag and 0.1 mag, respectively.
These are likely underestimates of the total extinction because the blue
wings are almost certainly affected by extinction as well, as discussed
above. From a study of spiral galaxy disks, Xilouris et al. (1999) found
that the central extinctions ranged from $A_V =$ 0.3 mag to 0.8 mag. In a
study of nuclear dust spirals, Martini \& Pogge (1999) determined color
excesses between arm and interarm regions that correspond to extinctions in
the range $A_V =$ 0.4 mag to 1.4 mag, assuming a standard Galactic
reddening curve (Savage \& Mathis 1979). Our STIS spectra yield reddening
values from the He~II $\lambda$1640/$\lambda$4686 line ratio that
correspond to $A_V =$ 0.3 mag to 1.5 mag (Kraemer \& Crenshaw 2000; Kraemer
et al. 2000). Thus, based on the above simple calculation, there is
sufficient dust in the inner galactic disk to account for the differences
in observed flux between the red and blue wings of the profiles.

Are there cases where a combination of outflows and extinction can lead to
a redshifted emission line? The answer is yes. In Figure 7, we show a test
case, based on the parameters listed in Table 2. In this extreme case, the
position angle and inclination of the galactic disk are such that the
blueshifted cone in the south is entirely occulted by the disk and the
redshifted cone in the north is not. To investigate this issue further, we
ran a simulation for every possible combination of disk inclination, bicone
inclination, and difference in position angle in 1\deg\ intervals, weighted
by the probability of observing a disk or bicone at a particular
inclination ($\propto$ sin $i$). Assuming a random distribution of these
parameters, we find that the percentage of the total population that shows
more extinction of the redshifted portion of the bicone than the
blueshifted portion is 17.2\%, 16.7\%, 15.6\%, and 13.7\% for half-opening
angles of the bicone of 30\deg, 40\deg, 50\deg, and 60\deg, respectively.
Thus, this model can explain the relatively small number of Seyfert
galaxies that show redshifted emission lines. However, the above
percentages cannot be directly compared to the observed values because the
observational errors do not allow us to determine the fraction of Seyferts
that have redshifted velocities between 0 and 50 km s$^{-1}$. Ideally, one
could attempt to match the distribution in Figure 1 by generating emission
line profiles from more realistic models that include thick disks and
bicones, a proper treatment of the dust extinction as a function of
position, realistic velocities and dispersions as a function of position,
and a more sophisticated error model, but that type of simulation is
beyond the scope of this paper.

\section{Summary and Conclusions}

We find that $\sim$40\% of Seyfert galaxies show radial velocity offsets
$|\Delta v| \geq$ 50 km s$^{-1}$ between their narrow emission lines and
host galaxy lines. The distribution of offsets peaks near zero km s$^{-1}$,
but has a strong blue tail extending up to $\sim$ $-$250 km s$^{-1}$; only
$\sim$6\% of the Seyferts in the sample show significantly redshifted
emission lines, up to 100 km s$^{-1}$. The Seyfert 1 distribution is
shifted to higher blueshifts on average.

To investigate the nature of the velocity offsets, we relied on our
previous investigations into the reddening, physical conditions, and
kinematics of the NLRs in NGC~1068 (Seyfert 2) and NGC~4151 (Seyfert 1),
which were based on spatially resolved {\it HST}/STIS spectra and
phototionization and kinematic models. The parameters from these studies
allowed us to generate geometric models of the biconical outflow in the NLR
and extinction by dust in the inner galactic disk. We also collapsed the
high-resolution {\it HST}/STIS spectra obtained at multiple slit locations
to simulate ground-based observations. The resulting [O~III] emission-line
profiles are asymmetric to the blue and have blue-shifted centroids, which
can be attributed to preferential extinction of the redshifted clouds in
the geometric models of these two Seyferts.

It is clear from the above examples that models that incorporate biconical
outflow and extinction by the inner galactic disk are more likely to
produce blueshifted velocity centroids than redshifted ones, because the
blueshifted clouds are closer to the observer and less likely to be
extincted in most cases. However, we have shown that redshifted velocity
centroids are possible in certain cases, for example when the disk is
highly inclined and covers most of the near cone. Furthermore, one
would expect from these models that more Seyfert 1s would show blueshifts
than Seyfert 2s, because the near cone would be pointed more directly at us
and therefore less extincted on average. All of these features of the
geometric models are in agreement with the observed velocity offsets.
However, we note that the [O~III] emission in NGC~1068 shows a higher
blueshift than that in NGC~4151, which indicates that individual variations
in velocities and extinctions of the NLR clouds amongst AGN are important
in determining the overall distribution in velocity offsets.

We conclude that the velocity offsets of emission lines in low-$z$ Seyfert
galaxies are primarily due to a combination of radial outflow in their NLRs
and extinction by dust on the same size scale as their NLRs. More exotic
explanations are not required. Furthermore, we find that the dust
responsible for the extinction must be primarily exterior to the ionized
gas. The dust features seen in the inner kpc of Seyfert galaxies, mostly in
the form of dust spirals (Martini et al. 2003, and references therein), are
the likely culprits.

As we have already noted, the bumps and peaks in the emission-line profiles
are due to large bright knots of emission-line gas traveling at their own
peculiar velocities, on top of the general outflow pattern. Thus, another
contributor to the velocity offsets could be an uneven distribution of
knots in velocity space in an AGN, resulting in more (or brighter)
blueshifted knots than redshifted ones, or vice-versa. This effect alone,
however, will not lead to an asymmetric distribution of velocity offsets.

It is instructive to compare the distributions of velocity offsets between
narrow emission and stellar absorption lines in the low $z$ ($<$ 0.043,
Nelson \& Whittle 1995) and moderate $z$ (0.34 $< z <$ 0.92, Comerford et
al. 2009) samples. The magnitudes of the velocity offsets in the two
distributions are essentially the same, and they have a similar percentage
of ``significant'' ($\gtsim$~50 km s$^{-1}$) offsets (41\% at low $z$, 35\%
at moderate $z$). However, the distribution for the moderate $z$ sample is
roughly symmetric around zero km s$^{-1}$, which is surprising, because
these AGN are likely to have outflows in their NLRs as well. In
fact, Comerford et al. use the symmetric distribution as a reason to
disfavor the outflows plus dust explanation and favor the inspiralling
black hole scenario. This, in turn, leads them to a merger fraction of
$\sim$30\% for red galaxies at these redshifts, which is a surprising large
number, particularly since these galaxies show no evidence for star
formation.

The nature of the Comerford et al. (2009) sample provides a couple of
possible explanations for a symmetric distribution of velocity offsets
in the context of mass outflows. Their sample contains only Seyfert 2
galaxies that have been selected to have red host galaxies, to specifically
avoid contamination from ionized gas around star-forming regions. The lack
of significant star formation suggests a lack of dusty gas, and therefore a
lack of extinction, in their NLRs. The velocity offsets would therefore be
dominated by uneven distributions of emission-line knots in at least some
of the AGN, with no preference for blueshifted profiles. Another possible
explanation is that the red colors are due to highly inclined disk
galaxies, which, as shown in Figure 7, will lead to a significant number of
redshifts and a more symmetric distribution of velocity offsets.
Nevertheless, there is a local example of velocity (and
spatial) offsets due to multiple AGN in the nearby galaxy NGC~3341 (Barth
et al. 2008), and one might expect more of these at higher redshifts from
galaxy mergers. However, the emission-line ratios in NGC~3341 indicate
significant star formation, as might be expected from mergers, and if it
was at the appropriate redshift, NGC~3341 would have not been included in
the Comerford et al. sample. Both the low-$z$ and moderate-$z$ samples are
rather small; larger sample would be helpful in clarifying the role that
mergers might play in producing velocity offsets.

Finally, we note that ground-based spectra of local Seyfert galaxies,
particularly those obtained at high spectral resolution (Vrtilek \&
Carleton 1985; Veilleux 1991b; Nelson \& Whittle 1995), show a wide variety
of narrow emission-line profiles. The profiles are often very irregular in
appearance, with bumps or shelves in their red or blue wings, and a number
of them show multiple peaks, like those in NGC~1068 (Figure 4), due to the
contributions of distinct emission-line knots. Of particular interest is
the [O~III] profile of Mrk~78 (Vrtilek \& Carleton 1985; Whittle et al.
1988; Nelson \& Whittle 1995), which shows two large well-defined peaks
separated by $\sim$800 km s$^{-1}$, and yet there is no evidence for
displaced or double SMBHs in Mrk~78. This velocity separation is actually
larger than those in the two ``dual AGN'' of Comerford et al. (2009), which
show velocity separations between double-peak profiles of 630 km s$^{-1}$
and 440 km s$^{-1}$. Thus, even these double-peaked AGN may not require
the presence of double SMBHs.

\acknowledgments

This research has made use of the NASA/IPAC Extragalactic Database (NED)
which is operated by the Jet Propulsion Laboratory, California Institute of
Technology, under contract with the National Aeronautics and Space
Administration. This research has made use of NASA's Astrophysics Data
System Bibliographic Services.

\begin{deluxetable}{cccccc}
\tablecolumns{6}
\footnotesize
\tablecaption{Heliocentric Radial Velocities and Offsets (in km s$^{-1}$)}
\tablewidth{0pt}
\tablehead{
\colhead{H~I 21-cm} & \colhead{Stellar} &
\multicolumn{2}{c}{[O~III] High Resolution} &
\multicolumn{2}{c}{[O~III] Low Resolution}\\
\colhead{Emission} & \colhead{Absorption} & 
\colhead{Centroid} & \colhead{Peak}& 
\colhead{Centroid} & \colhead{Peak}
}
\startdata
\multicolumn{6}{c}{NGC~1068}\\
1137$\pm$3$^a$ & 1149$\pm$8$^c$ & 974$\pm$7 & 1128$\pm$5 &
962$\pm$8 & 988$\pm$7\\
$\Delta v$ &12$\pm$ 9 &$-$163$\pm$8 & $-$9$\pm$6 &$-$175$\pm$9
&$-$149$\pm$8\\
\hline
\multicolumn{6}{c}{}\\
\multicolumn{6}{c}{NGC~4151}\\
995$\pm$3$^b$ & 1006$\pm$18$^c$ & 935$\pm$5 & 998$\pm$4 &
926$\pm$3 & 981$\pm$4\\
$\Delta v$ &11$\pm$18 &$-$60$\pm$6 & 3$\pm$5 &$-$69$\pm$4 &$-$14$\pm$5
\enddata
\tablenotetext{a}{Bottenelli et al. 1990}
\tablenotetext{b}{de Vaucouleurs et al. 1991}
\tablenotetext{c}{Nelson \& Whittle 1995}
\end{deluxetable}

\begin{deluxetable}{llll}
\tablecolumns{4}
\footnotesize
\tablecaption{NLR Bicone and Host Galaxy Parameters$^a$}
\tablewidth{0pt}
\tablehead{
\colhead{Parameter} & \colhead{NGC~1068} &
\colhead{NGC~4151} & \colhead{Test}
}
\startdata
P.A. (Galaxy)    &106\deg      &22\deg       &90\deg \\
Incl. (Galaxy)   &40\deg (SW)  &20\deg (W)   &60\deg (S) \\
P.A. (Bicone)    &30\deg       &60\deg       &0\deg \\
Inclin. (Bicone) &5\deg (NE)   &45\deg (SW)  &30\deg (S) \\
H.O.A. (Bicone)  &40\deg       &33\deg       &30\deg  \\
\enddata
\tablenotetext{a}{ See the text for a description of the parameters,
obtained from Das et al. (2005, 2006) for NGC~1068 and NGC~4151. The
letters in parentheses indicates the side closest to us.}
\end{deluxetable}

\clearpage

\figcaption[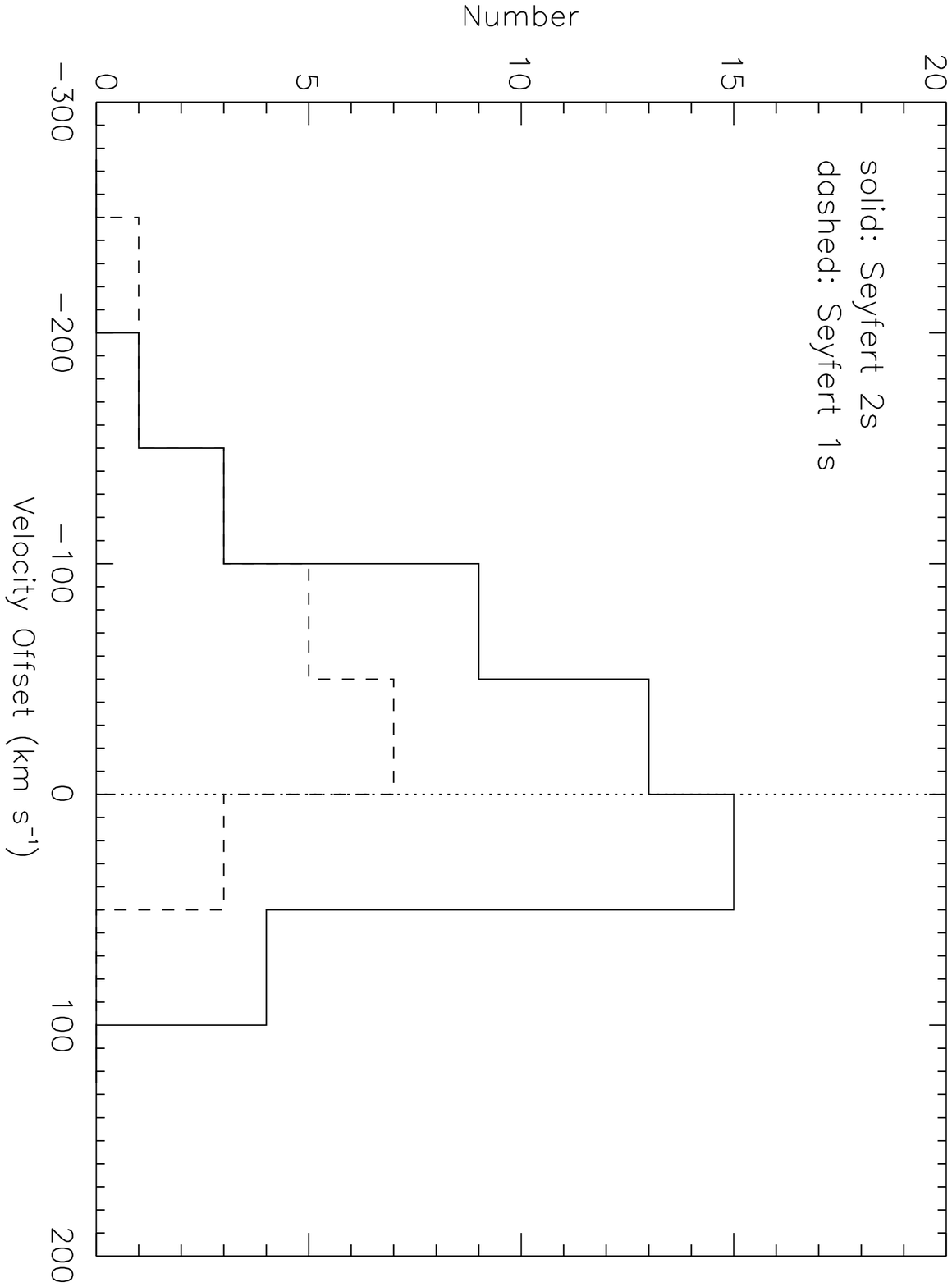]{Histograms of the velocity offsets between emission
lines and absorption lines ($\Delta v = v_{emis} - v_{abs}$) in Seyfert
1 and Seyfert 2 galaxies, from the measurements of Nelson \& Whittle
(1995).}

\figcaption[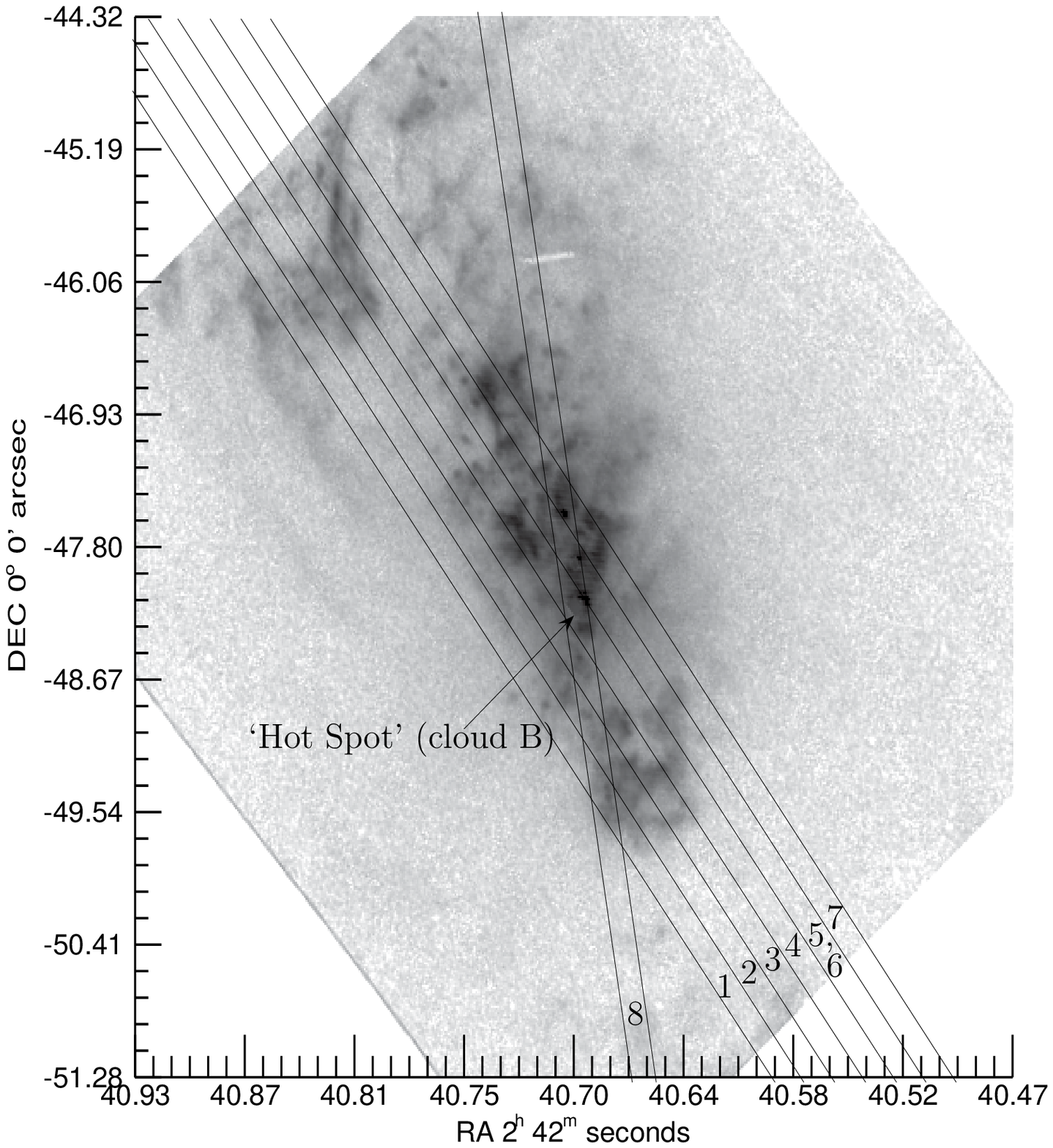]{[O~III] emission-line image of the NLR in NGC 1068,
obtained with the {\it HST} Faint Object Camera, showing the slit positions
for the STIS long-slit observations with the G430M grating. The SMBH is
0\arcsecpoint15 south of the hot spot (Das et al. 2006).}

\figcaption[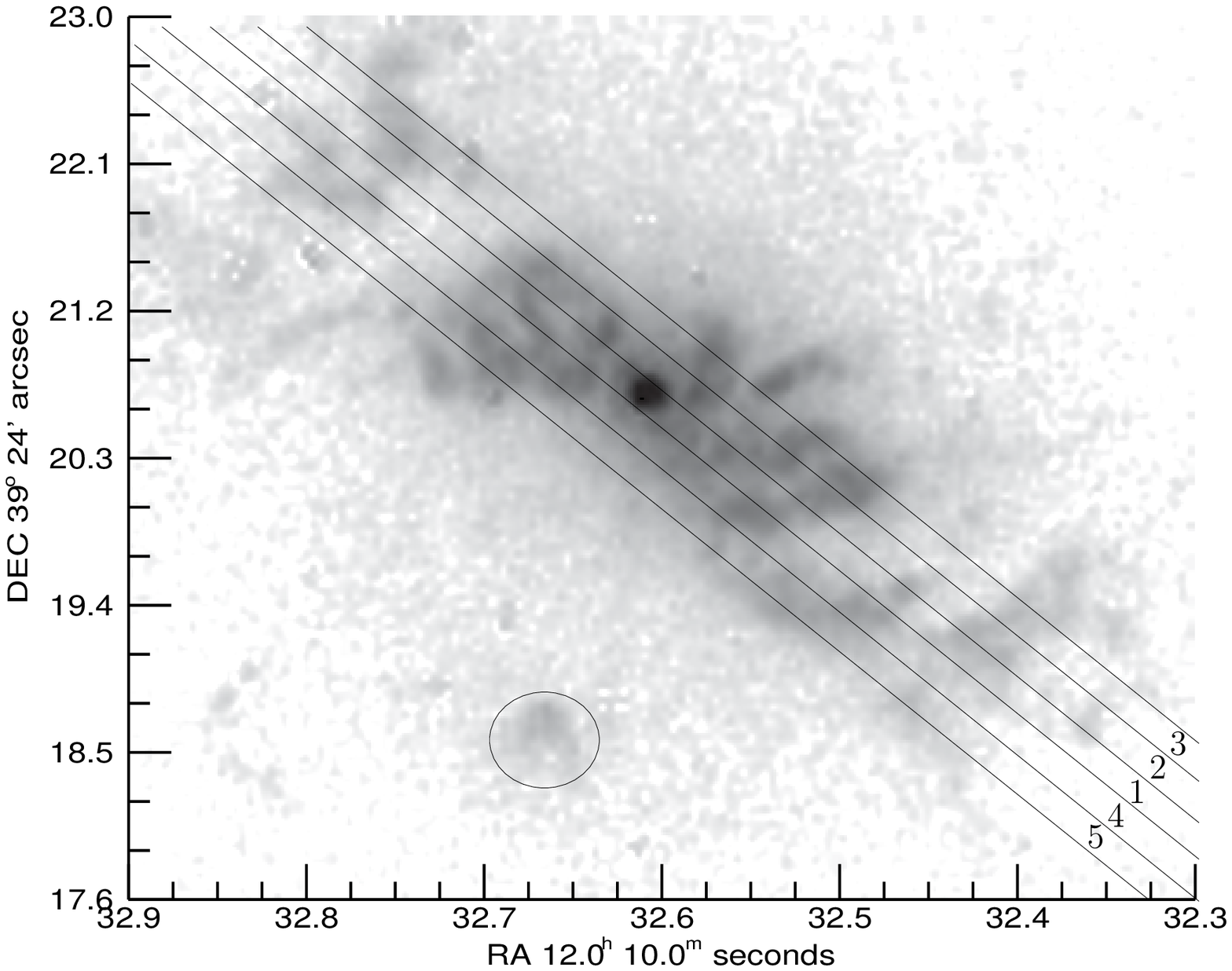]{[O~III] emission-line image of the NLR in NGC 4151,
obtained with the {\it HST} Wide Field Planetary Camera 2, show the slit
positions for the STIS G430M long-slit observations (see also Das et al.
2005). The circled feature is due to a reflection in the camera.}

\figcaption[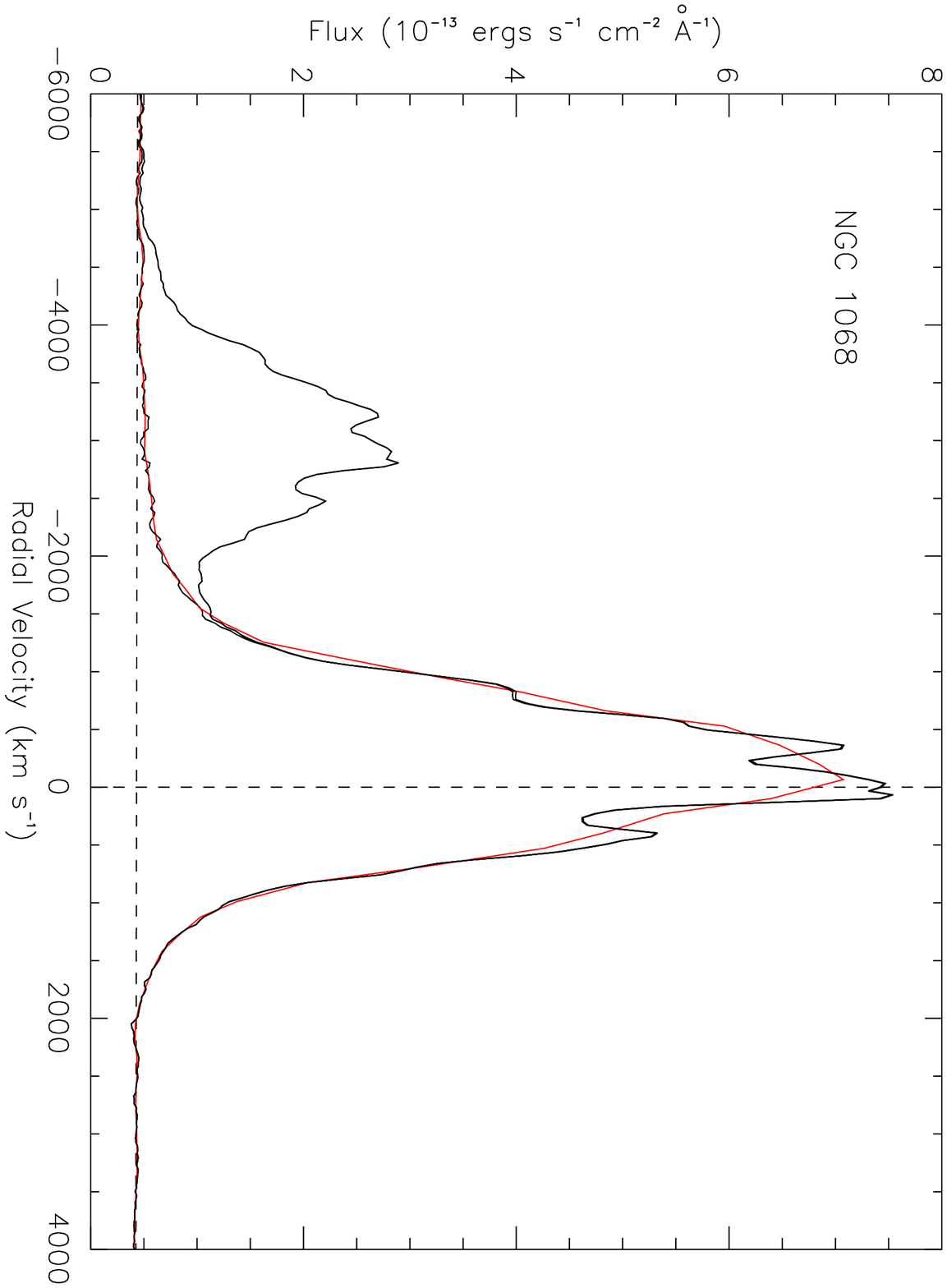]{Collapsed spectra of the [O~III] emission lines in
NGC~1068, obtained by adding together all of the spectra from the STIS
long-slit observations and plotting the flux vs. radial velocity, where
zero km s$^{-1}$ is defined by H~I 21-cm emission. The solid black lines
shows the unsmoothed (R $\approx$ 9000) spectrum, and the deblended blue
wing of [O~III] $\lambda$5007 underneath the [O~III] $\lambda$4959 line.
The solid red line shows the heavily smoothed (R $\approx$ 1000) version of
the [O~III] $\lambda$5007 line.}

\figcaption[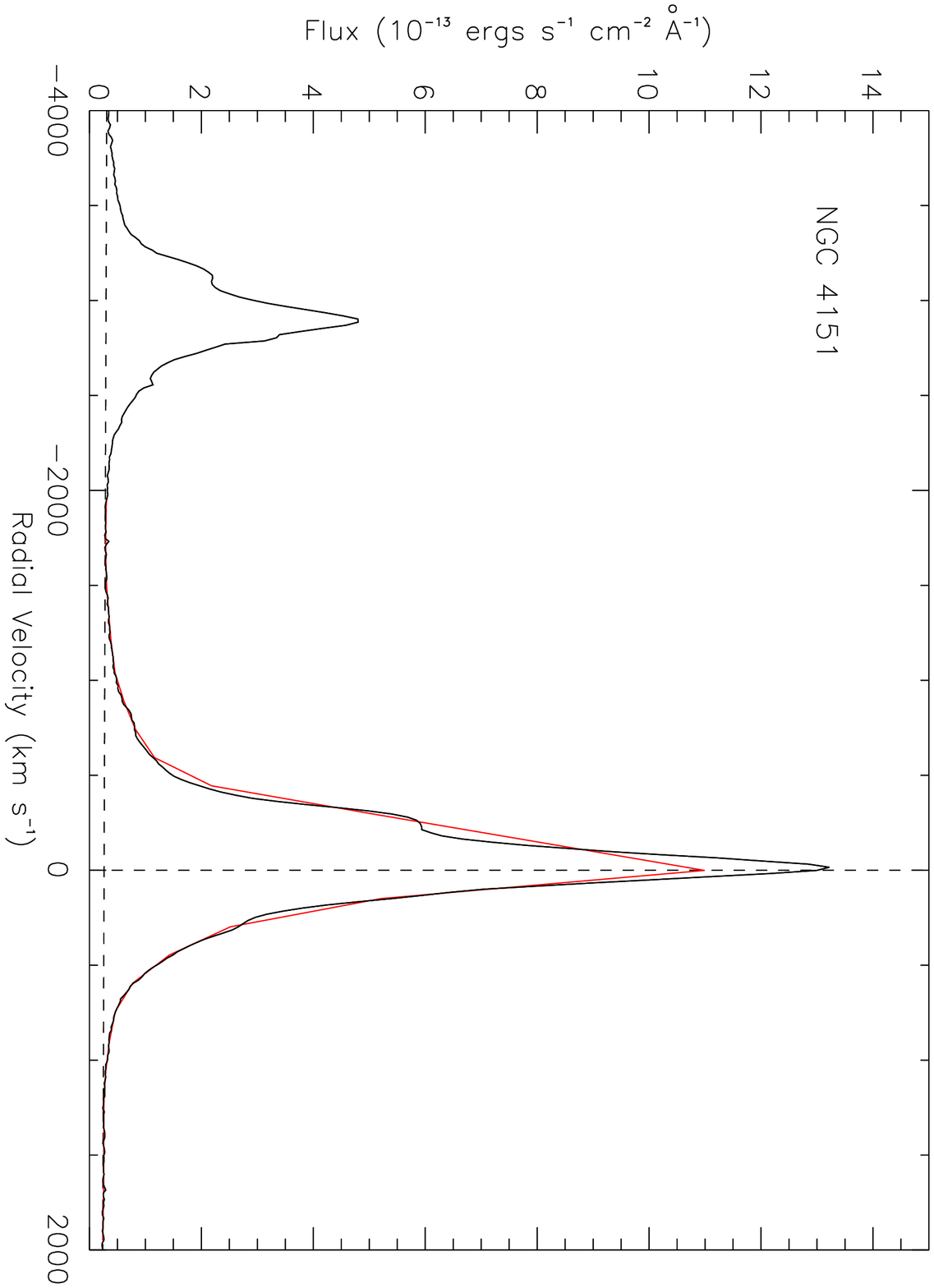]{Collapsed spectra of the [O~III] emission lines in
NGC~4151, obtained by adding together all of the spectra from the STIS
long-slit observations and plotting the flux vs. radial velocity, where
zero km s$^{-1}$ is defined by H~I 21-cm emission. The solid black line
shows the unsmoothed (R $\approx$ 9000) spectrum. The solid red line shows
the heavily smoothed (R $\approx$ 1000) version of the [O~III]
$\lambda\lambda$4959, 5007 lines.}

\figcaption[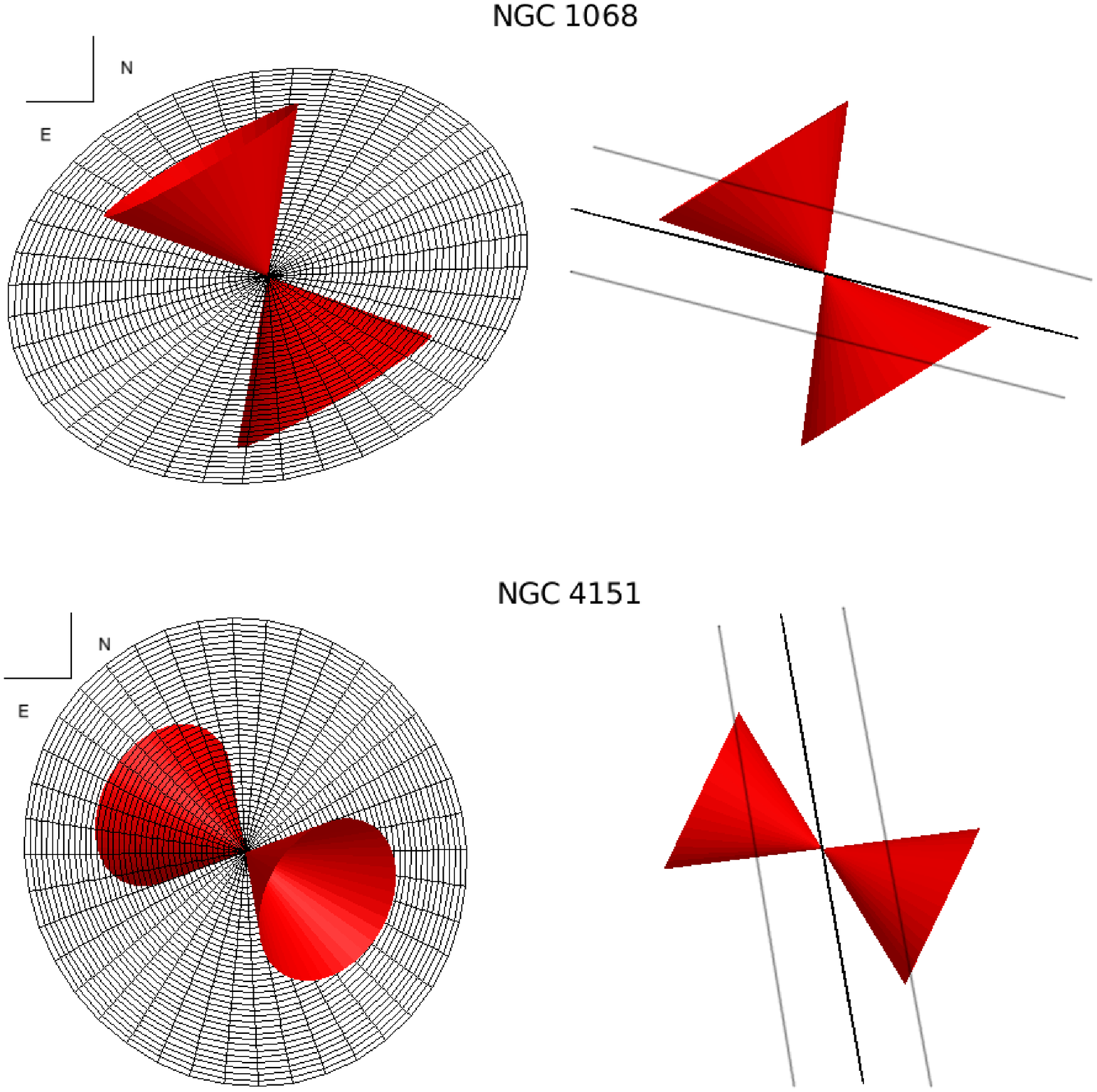]{Geometric models of the NLR bicones and the inner
galactic disks in NGC~1068 and NGC~4151, based on the parameters in Table
2. The left-hand side shows our view, and the right hand side shows a view
in which the bicone axis is in the plane of the sky and the galactic disk
is edge-on (our view is to the upper right and above the page for
each). The thin gray lines represent a disk scale height of 200 pc for
a bicone that is 800 pc in length along its axis.}

\figcaption[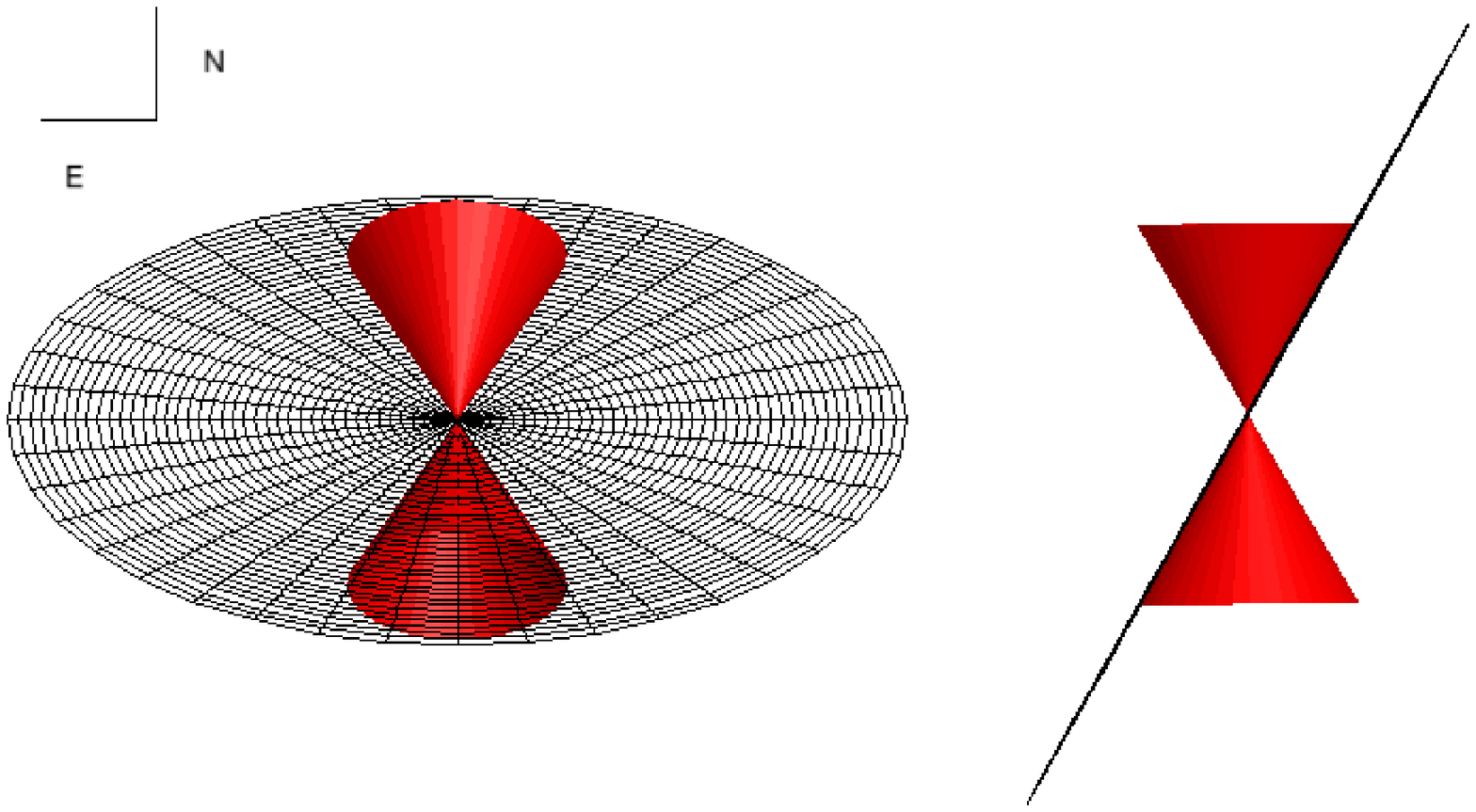]{Test model of a hypothetical Seyfert galaxy with the
NLR and disk parameters given in Table 2. The left-hand side shows our
view, and the right hand side shows a view in which the bicone axis is in
the plane of the sky and the galactic disk is edge-on (our view is
to the lower left and in the page).}

\clearpage
\begin{figure}
\plotone{f1.eps}
\\Fig.~1.
\end{figure}

\clearpage
\begin{figure}
\plotone{f2.eps}
\\Fig.~2.
\end{figure}

\clearpage
\begin{figure}
\plotone{f3.eps}
\\Fig.~3.
\end{figure}

\clearpage
\begin{figure}
\plotone{f4.eps}
\\Fig.~4.
\end{figure}

\clearpage
\begin{figure}
\plotone{f5.eps}
\\Fig.~5.
\end{figure}

\clearpage
\begin{figure}
\plotone{f6.eps}
\\Fig.~6.
\end{figure}

\clearpage
\begin{figure}
\plotone{f7.eps}
\\Fig.~7.
\end{figure}

\end{document}